\documentstyle[color,epsfig,psfig]{article}
\input{epsf}
\input{prepictex}
\input{pictex}
\input{postpictex}

\newcommand{\true}{\mbox{\em true}}
\newcommand{\false}{\mbox{\em false}}

\newcommand{\hst}{\\\hspace*{4mm}}
\newcommand{\hstt}{\\\hspace*{8mm}}
\newcommand{\hsttt}{\\\hspace*{12mm}}

\newcommand{\pf}{\noindent\mbox{\bf Proof : }}

\def\qed{\ifmmode\|\else{\unskip\nobreak\hfil
\penalty50\hskip1em\null\nobreak\hfil$\|$
\parfillskip=0pt\finalhyphendemerits=0\endgraf}\fi}

\newtheorem{theorem}{THEOREM}
\newtheorem{lemma}[theorem]{LEMMA}

\newtheorem{definition}{Definition}

\newenvironment{list1}{\begin{list}{$\bullet$}
{\topsep 0 pt \parsep 0 pt \partopsep 0 pt \itemsep 0 pt}}{\end{list}}
\newenvironment{list2}{\begin{list}{$-$}
{\topsep 0 pt \parsep 0 pt \partopsep 0 pt \itemsep 0 pt}}{\end{list}}

\newcounter{cabbage1}

\newcounter{cabbage2}

\newcounter{cabbage3}

\newcounter{bean1}

\newcounter{bean2}

\newcounter{bean3}

\newcounter{bean4}

\newcounter{bean5}

\newcounter{bean6}

\title{Numerical Coverage Estimation for the Symbolic Simulation
of Real-Time Systems
\thanks{The
work is partially supported by NSC, Taiwan, ROC under
grants NSC 90-2213-E-001-006, NSC 90-2213-E-001-035, and the by the Broadband network protocol verification
project of Institute of Applied Science \& Engineering Research, Academia Sinica, 2001.
}}
\author{Farn Wang\\
Dept. of Electrical Engineering, National Taiwan University\\
Taipei, Taiwan 106, Republic of China\\
+886-2-23635251 ext. 435; FAX
+886-2-23671909; \verb+farn@cc.ee.ntu.edu.tw+\\ Tools available
at: \verb+http://val.iis.sinica.edu.tw/+\\\\[2mm]
Geng-Dian Hwang, Fang Yu\\
Institute of Information Science, Academia Sinica\\ Taipei, Taiwan
115, Republic of China
        }

\date{}
\begin{document}
\baselineskip 13pt
\maketitle
\thispagestyle{empty}
\pagestyle{plain}

\begin{abstract}
Three numerical coverage metrics for the symbolic simulation of dense-time systems and
their estimation methods are presented.
Special techniques to derive numerical estimations of dense-time state-spaces
have also been developed.
Properties of the metrics are also discussed with respect to four criteria.
Implementation and experiments are then reported.
\end{abstract}

\noindent {\bf Keywords:}
coverage, verification, symbolic simulation, real-time

\section{Introduction}

Presently, with verification and integration costs increasing to more than
50 percent of the development budget in real-world projects, it is
more and more difficult to use traditional simulation technology
to acquire enough trace coverage to confidently create
system designs. 
As well, application of the new
formal verification technology is still hampered by its 
intrinsic complexity.   
In the forseeable future, we expect that 
simulation and formal verification will be combined for 
verification of large-scale real-time systems. 
Symbolic simulation is such a combination\cite{SB95}. 
It uses symbolic techniques\cite{BCMDH90,Bry86,Dill89,HNSY92} 
of formal verification to represent space of simulation traces so that
abstract (as opossed to concrete) behaviors can be observed in a trace. 
For verification of real-time systems, tools like
UPPAAL\cite{PL00} and RED\cite{Wang00,Wang01,Wang03,WHY03} support
symbolic simulations.

Current symbolic simulation technology for real-time systems
is still not as developed as that for untimed systems 
like Very Large Scale Integration (VLSI) Systems. 
For one thing, the important concept of {\em coverage} can
be used to both estimate the value of a set of traces and to direct
the generation of new traces. In short, coverage is how much has been verified of
the target to be verified. 
The importance of this concept is that, in real-world
projects, it is usually the case that we do not have enough
resources to either run enough traces to obtain 
confidence, or to complete formal verification tasks.
Product designs usually need to be released before we can obtain 100\%
confidence in the designs.
Therefore it is important that we have some type of metric to evaluate confidence in our designs. 
A common coverage metric for simulation is {\em code coverage},
which measures the proportion of already-executed Hardware Description
Language (HDL) statements during simulation. 
{\em State and transition coverages} are used in control state machines\cite{YDJ95}. 
These coverage metrics have proven to be effective in bug escape reduction by pointing
out coverage holes in the test suite. 
Coverage goals are used to
measure the degree of confidence in the total verification effort,
and to help the design team predict the optimal time for design
release\cite{Beizer90}.

The coverage metrics used in traditional simulation, which is based on concrete traces,
may not be directly applicable to the symbolic simulation of dense real-time systems, since
there are infinitely many concrete traces and states for such systems.
For example, the {\em visited-state} coverage metric\cite{BF01,RPS01}, which uses concrete reachable states
in FSM to estimate coverage, is not suitable for the symbolic simulation of dense real-time systems.
If we directly apply this coverage metric to the simulation of dense real-time systems, 
we will always have 0\% coverage since the ratio of finitely many concrete states over the infinite 
reachable state-space is always zero.

To this end, we propose techniques to estimate numerical coverage for the symbolic simulation of dense real-time systems. 
As mentioned above, the states in such systems are dense, and the question follows that how do we count the states, since 
"dense" stands for uncountable. 
Nevertheless, our techniques can estimate the covered proportion of the reachable state-space, and it is efficient and meanful.
We approach the question by adopting the region-relation\cite{ACD90} to partition dense state-space, and 
propose a method to estimate the size of each proportion in section~\ref{sec.rcm.zone}.
We believe that our techniques can be used to help future
development of various coverage-based verification techniques -
including the design of new coverage metrics and coverage-based
test-pattern generation - in real-time systems.

Before we can estimate numerical coverage, we must design a
metric and its estimation procedure. 
This engenders the first
research issue of this work, that is, how do we know if a metric
is good ?   
In section~\ref{sec.criteria}, we present four criteria
to serve as guidelines in coverage metric design.  
These criteria are: 
{\bf accountability} (each basic {\em portion} of the {\em target function} is
counted once and only once), {\bf coverability} (100\% coverage
estimation is achievable),
{\bf efficiency} (the overhead in computing the coverage
estimation is low), and {\bf discernment} (risk states are discernable). 
According to these criteria, we adapt three
coverage metrics from traditional testing research and development
techniques to implement them in real-time systems. 
These three
new metrics are: {\em timed automata arc coverage metric(ACM)}, 
{\em back-and-forth region coverage metric(RCM)}\cite{RPS01}, and
{\em triggering condition coverage metric(TCM)}.
ACM is a straightforward adaptation from the technology of VLSI simulation, whereas RCM
and TCM are not. 
For dense-time systems, RCM and TCM are more precise in the estimation
of coverage by considering state-space coverage. 
We shall prove in a lemma that RCM has enough power to
discern reachable risk states while ACM lacks sufficient power.
To maintain the four criteria
for dense-time state-spaces of real-time systems, we have
developed techniques to quantitatively estimate the volume of a
state-space, and to significantly prune irrelevant state-space portions
from a verification task.

To demonstrate the usefulness of our techniques for real-world
projects, in section~\ref{sec.experiments}, we have modeled and verified
the {\em  Logical Link Control and
Adaptation Layer Protocol(L2CAP)} of Bluetooth
specification\cite{Haartsen01}.
Bluetooth is a widely adopted wireless communication standard in the industry.
We model two devices that communicate with the Bluetooth L2CAP and carry out simulation
experiments to gather data on numerical coverage estimation.
We then compare the three coverage metrics with respect to our
experiment data and the four criteria.

In section~\ref{sec.relwork}, we review related works.
In section~\ref{sec.sysmodel}, we briefly present our verification framework with
timed automata.
In section~\ref{sec.criteria}, we discuss the four criteria for effective coverage metrics.
In sections~\ref{sec.acm} through \ref{sec.tcm},
we present our three coverage metrics and their estimation procedures.
Finally, in section~\ref{sec.experiments},
we report on our experiment results with the Bluetooth L2CAP
and discuss the implications of the experiment data.

\section{Previous work \label{sec.relwork}}

Coverage techniques have been widely discussed and applied in testing, 
simulation and formal verification of various system designs.
In software testing, people use {\em control flowgraphs} \cite{Beizer90}, which are composed of processes,
decisions, and junctions. 
Given a set of program stimuli, one can
determine the statements activated by the stimuli with the
coverage metrics of the flowgraphs. 
Programming code metrics measure syntactical characteristics of a code w.r.t its execution
stimuli. For example, {\em line coverage metric} measures the
number of distinct statements visited during the course of
execution, {\em branch coverage} measures the number of distinct branch decisions, and 
{\em path coverage} measures the number of distinct paths 
 (i.e. a unique combination of branch decisions and statements) 
exercised due to its execution stimuli\cite{BF01}.
The number of paths in a program may be exponentially related to program size which greatly
hinders attaining 100\% path coverage in software testing.

Coverage analysis techniques proposed for general HDL programs
include: guaranteed coverage of every statement\cite{CK93},
transition coverage of a test set\cite{HMA95}, and abstraction of
models and semantic control over transition
coverage\cite{GFLLUW96}. 
Fallah provides OCCOM\cite{FDK98} to address the observability issue. 
Most of these HDL metrics are used to drive test-generation in simulation analysis.

Ho et al.\cite{HKHZ00} proposed a coverage metric to estimate the
"completeness" of a set of properties verified in model-checking
FSM against a subset of CTL.
A symbolic algorithm is also presented.
Chockler et al.\cite{CKV01} also suggested several coverage metrics to measure 
completeness of a verified specification, and to find uncovered parts.

Dill proposed a way to bridge the
gap between simulation and formal verification\cite{Dill98}.
Generator of Test Cases for Hardware Architecture(GOTCHA) is a
prototype coverage-driven test generator implemented as an
extension to the Mur$\phi$ model-checker\cite{MDAYGR99}. 
It supports state and transition coverage analysis in FSM. 
On completion of the entire reachable state-space enumeration,
a random coverage task is chosen from those not yet covered.

Opposed to previous works with untimed or discrete-time systems, we apply coverage techniques
in our symbolic simulator with dense-time model.
One difficulty arises in the design of
meaningful metrics to estimate state-spaces which are both dense
and infinite.
Traditional state and transition coverage metrics
for untimed or discrete-time systems cannot be directly copied
since metrics may always be zero based on the dense domain.

\section{Framework of verification \label{sec.sysmodel}}

\subsection{System model}

We use the widely accepted model of {\em timed automata (TA)}\cite{ACD90}.
As we assume familiarity with this model, we will not go into much detail.
A TA is a finite-state automaton equipped with
a finite set of clocks which can hold nonnegative real-values.
A TA can stay in only one {\em mode} (or {\em control location}) at a time.
In operation, one transition can be triggered when its corresponding triggering
condition is satisfied.
Upon being triggered, the TA instantaneously transits from one
mode to another and resets some clocks to zero.
Between transitions, all clocks increase their readings at a uniform
rate.

For convenience, given a set $Q$ of modes and a set $X$ of clocks, we use $B(Q,X)$ as
the set of all Boolean combinations of mode predicate $\mbox{\tt mode}=q$,
where {\tt mode} is a special auxiliary variable, and
inequalities of the forms $x-x'\sim c$, where
$q\in Q$, $x,x'\in X\cup\{0\}$, ``$\sim$'' is one of
$\leq, <,=,>,\geq$, and $c$ is an integer constant.

{\definition \underline{\bf timed automata (TA):}}
A timed automaton $A$ is given as
a tuple $\langle X, Q, I, \mu, T, \tau, \pi\rangle$
with the following restrictions:
$X$ is the set of clocks,
$Q$ is the set of  modes,
$I\in B(Q,X)$ is the initial condition on clocks,
$\mu:Q\mapsto B(\emptyset,X)$ defines the invariance condition of each mode,
$T\subseteq Q\times Q$ is the set of transitions, and
$\tau:T\mapsto B(\emptyset,X)$ and $\pi:T\mapsto 2^X$ respectively
define the triggering condition and the clock set to reset
of each transition.
\qed
\vspace*{2mm}

A {\em valuation} of a set is a mapping from that set to another set.
Given an $\eta\in B(Q,X)$ and a valuation $\nu$ of $X$, we say
$\nu$ {\em satisfies} $\eta$, in symbols $\nu\models\eta$,
iff $\eta$ will be evaluated $\true$ when the variables in $\eta$ are
interpreted according to $\nu$.

{\definition \underline{\bf states:}}
A state $\nu$ of $A=\langle X, Q, I, \mu, T, \tau, \pi\rangle$ is a valuation of
$X\cup\{\mbox{\tt mode}\}$ such that
\begin{list1}
\item $\nu(\mbox{\tt mode})\in Q$ is the mode of
    $A$ in $\nu$; and
\item for each $x\in X$, $\nu(x)\in {\cal R}^+$
    such that ${\cal R}^+$ is the set of nonnegative real numbers
    and $\nu\models\mu(\nu(\mbox{\tt mode}))$.
\qed
\end{list1}
\vspace*{2mm}
For any $t\in {\cal R}^+$, $\nu+t$ is a state identical to $\nu$
except that for every clock $x\in X$, $\nu(x)+t = (\nu+t)(x)$.
Given $\bar{X}\subseteq X$, $\nu \bar{X}$ is a new state identical
to $\nu$ except that for every $x\in \bar{X}$, $\nu\bar{X}(x)=0$.

{\definition \underline{\bf runs:}}
Given a timed automaton $A=\langle X, Q, I, \mu, T, \tau, \pi\rangle$,
a {\em run} is an infinite sequence of state-time pairs
$(\nu_0,t_0)(\nu_1,t_1)\ldots(\nu_k,t_k)\ldots$
such that $\nu_0\models I$ and $t_0 t_1 \ldots t_k\ldots$
is a monotonically increasing real-number (time)
divergent sequence,
and for all $k\geq 0$,
\begin{list1}
\item for all $t\in [0, t_{k+1}- t_k]$,
    $\nu_k+t\models\mu(\nu_k(\mbox{\tt mode}))$; and
\item either $\nu_k(\mbox{\tt mode})=\nu_{k+1}(\mbox{\tt mode})$
    and $\nu_k+(t_{k+1}-t_k) = \nu_{k+1}$; or
    \begin{list2}
    \item $(\nu_k(\mbox{\tt mode}),\nu_{k+1}(\mbox{\tt mode}))\in T$ and
    \item $\nu_k+(t_{k+1}-t_k)\models\tau(\nu_k(\mbox{\tt mode}),
                \nu_{k+1}(\mbox{\tt mode}))$ and
    \item $(\nu_k+(t_{k+1}-t_k))
        \pi(\nu_k(\mbox{\tt mode}),\nu_{k+1}(\mbox{\tt mode}))
        =\nu_{k+1}$.
\qed
    \end{list2}
\end{list1}

\subsection{Procedure of simulation and coverage analysis}

We adopt the {\em safety-analysis problem} as our verification
framework for simplicity.
In this framework, we want to check whether an unsafe state can be reached by
repetitive generation of symbolic traces.
Formally speaking, a safety analysis problem instance 
consists of a timed automaton $A$ and a
{\em safety state-predicate} $\eta\in B(Q,X)$.
$A$ is {\em safe} with respect to $\eta$, in symbols $A\models \eta$,
iff for all runs $(\nu_0,t_0)(\nu_1,t_1)\ldots(\nu_k,t_k)\ldots\ldots$,
for all $k\geq 0$,
and for all $t\in [0, t_{k+1}- t_k]$, $\nu_k+t\models\eta$,
i.e., the safety requirement is guaranteed.

We construct our main procedure based on the well-discussed
symbolic procedure, called {\tt next}(), to compute a
symbolic post-condition after a discrete transition and
time-progress\cite{HNSY92,LPW95}.
Our symbolic simulation procedure takes
the following form (details on coverage estimation in statements (1) and (4) will be explained
in sections~\ref{sec.acm} through \ref{sec.tcm}).
\label{proc.ss} 
\vspace*{3mm} 
\hrule 
\vspace*{2mm}
\noindent {\tt Symbolic\_Simulate}$(A,\eta)$ /* $A$ is a TA; $\eta$ is the safety state predicate. */ \{
\hst Compute the numerical estimation $f$ of the whole {\em target function} $F$. \hfill (1)
\hst    Let $\phi:=I$; $\phi':=\false$; $v:=0$;
\hst	While (true) \{
\hstt       Let $\phi':=\phi;$
\hstt       Select a subspace $\psi$ of $\phi$ and a set $\bar{T}$ of transitions
\hsttt            (possibly based on the value $\phi$ and $\bar{T}$); \hfill (2)
\hstt       Compute $\phi:=\phi\vee\bigvee_{e\in \bar{T}}{\tt next}(A,\psi,e)$; \hfill (3)
\hstt       Compute the estimation $v$ of the verified proportion $V$
\hsttt		of the whole {\em target function}; \hfill (4)
\hstt       Print $v/f$ as the new numerical coverage estimation.  \hfill (5)
\hstt       If $\phi\wedge \neg\eta\neq \emptyset$,
\hsttt		print out "a risk state is reachable" and exit;
\hstt	    else if $v/f \geq threshold$,
\hsttt	    	print out "The threshold of the chosen coverage metric is reached" and exit;
\hstt	    else $\phi=\phi'$
\hsttt	    	print out "no risk states are found" and exit; \hfill (6)
\hst	\}
\\\}
\vspace*{2mm}
\hrule
\vspace*{3mm}

In this manuscript, we use the term {\em "portion"} to 
mean a basic unit of the target function in the 
estimation of trace coverage.  
Formally speaking, 
given a coverage metric, 
a portion is an equivalence class of (syntactic or semantic) entities of 
the target function in which two entities cannot be distinguished by 
the given coverage metric. 
The target function is conceptually defined as the set of all portions.  
Coverage means that how much of the target function has appeared 
in a set of simulation traces.  

In the case of line coverage, a portion is a statement line. 
For state-coverage, a portion is a concrete state of the verification target. 
We can also use regions as the portion in the simulation of dense-time systems.  
In this case, a portion can contain infinitely many concrete states. 

The {\em target function} can be the set of TA transitions (arcs), the regions of whole reachable state-space,
or the regions of the triggering-conditions of all transitions in this paper.

In statement (2), we allow for the selection flexibility of various search strategies.
Indeed, we have already implemented game-based, goal-oriented, and random strategies \cite{WHY03}.
A subspace $\psi$ of the verified state-space $\phi$
and a set $\bar{T}$ of selected transitions are fed to procedure {\tt next}()
to compute the new next-step state-space after transitions and time-progress in statement (3).
In statement (5), the coverage is numerically estimated as the ratio of the
already-verified proportion of the whole target function.
The infinite loop can continue until a risk state is reached,
or until we feel that enough confidence has been established
(coverage of the function has reached some specific $threshold$),
or until we have reached the fixed point and finished the exhaustive search  in statement (6).

However, our simulation framework is actually more general than
simulation.
For example, if in every iteration, we choose
$\psi=\phi$ and $\bar{T}=T$ in statement (2), the whole procedure becomes a forward
reachability analyzer.
In the next few sections, we will discuss
how to compute coverage estimations according to our three
coverage metrics.
As for the use of various strategies
to guide searches, we believe it deserves more effort in the
future.

\section{Criteria for good coverage metrics \label{sec.criteria}}

A good coverage estimation should tell us
what proportion of a target function has been covered.
We can partition a {\em target function} into {\em portions} and use an
estimation function $\epsilon()$(from the set of portions to the set of nonnegative reals)
to numerically estimate coverage.
Formally speaking,
$\epsilon:F\mapsto {\cal R}^+$ where ${\cal R}^+$ is the set of nonnegative reals.
The whole target function can then be estimated as $f=\sum_{p\in F}\epsilon(p)$,
and the current covered proportion $V$ of the target function, i.e. the verified subset of $F$,
can be estimated as $v=\sum_{p\in V}\epsilon(p)$.
In ACM, a portion represents a physical entity
(i.e. a transition) of the automata,
then coverage of that portion means that the physical entity has been
used in some traces.
In RCM and TCM, a portion represents a state subspace(i.e. a region),
then the occurrence of any state in the portion along some simulation traces
indicates the portion has been covered.

In practice, it can be difficult to design a good metric for
dense-time systems.
For example, we may want to use visited states as portions.
Then in a dense-time state-space, 
we have to decide how a state should correspond to a portion.
For discernment, a natural choice of a portion is the region presented in \cite{ACD90}.
But it is very expensive (PSPACE-complete) to compute a precise representation of
the entire reachable region set.
A naive solution to this challenge is to use symbolic techniques with
the popular data-structure of DBM (difference-bounded matrix)\cite{Dill89}.
The challenge is that DBMs are not necessarily disjointed from one another.
If we sum up portion estimations of each state-space using a DBM
to calculate the total estimation, it is likely that some portions will be counted
more than once.

After experimenting with various coverage metrics and their computation methods,
we have identified the following four criteria for effective numerical coverage metrics.
\begin{list1}
\item {\bf accountability:}
    This assures that each portion of the target function is accounted for 
    once and only once.
    If accountability is not maintained, we may run into the two following
    bizarre situations.
    First, some portions may not be accounted for and
    thus engineers simply cannot trust the metrics to check if all function portions have been
    covered.
    Second, it may happen that some portions are counted more than once
    and thus full coverage estimation is greater than 100\% which makes no sense at all.
    Thus, accountability is the most important criterion.
    If we are going to use state-space or its abstraction to estimate coverage 
    of target functions in dense-time systems, then we must develop new techniques,
    other than DBMs, to assure each portion is accounted for exactly once.
\item {\bf coverability:}
    This means that $\sum_{p\in V}\epsilon(p)=\sum_{p\in F}\epsilon(p)$ can be expected
    at the end of a symbolic simulation if enough traces have been generated.
    This is desirable in that 100\% coverage can be the goal for verification.
    Moreover, if engineers decide to stop verification at 80\% coverage, they
    can roughly estimate confidence in their products.
    It is likely to stump verification engineers if
    the coverage estimation converges at a small percentage number,
    no matter how many traces they have generated.
    One-hundred percent coverage can only be achieved if we have
    a precise numerical estimation of the entired target function $F$ to be verified.
\item {\bf efficiency:}
    This criterion measures the overhead in the computation
    of both the $f$ (at statement (1)) and the $v$ (at statement (4))
    in procedure {\tt Symbolic\_Simulate}().
    If complex formal reachability analysis is used to compute these
    two estimations, it is not worthwhile to estimate the coverage.
    In this work, we base our coverage estimation on transition-countings and
    state-space abstraction techniques and can efficiently calculate estimations
    in our three metrics.
\item {\bf discernment:}
    This criterion assesses the capability of a metric to
    discern risk states.
    This can be an issue when, in some metrics, risk states and non-risk states
    are likely to fall in the same $portion$.
    A metric that frequently fails to detect existing risk states at
    a high numerical coverage may give users unjustified and false confidence
    on their system designs.  
\end{list1}
The third and fourth criteria are kind of contradictory to each other.
In a lot of cases, in order to discern risk states,
we not only have to partition the portions intelligently,
but we also have to partition them in great resolution.
And this usually results in high complexity and low efficiency
to reach high coverage with enough traces through the huge space of portions.

In the following, we shall use these four criteria to evaluate the coverage metrics
presented in the next few sections.

\section{TA arc coverage metric (ACM)\label{sec.acm}}

This is a straightforward adaptation from the technology of VLSI
simulation and testing. 
In the computation of FSM arc coverage for
VLSI, we conceptually transform a circuit to a finite-state
automaton (FSM) and use the set of already-triggered transitions as
$V$ and the set of executable transitions as $F$ to compute
coverage estimation\cite{BF01,RPS01}.  
The same definition of FSM arc coverage can
readily be copied for the simulation of timed automata (TA). 
That is, we can also use the arcs of TAs to estimate coverage in the {\em TA arc coverage metric (ACM)}. 
The straightforwardness of this metric has many desirable features.
Each portion corresponds to an executable transition and the
estimation function $\epsilon_{ACM}()$ maps everything to 1. 
The numerical estimation $f$ of the whole target function in statement
(1) of procedure Symbolic\_Simulate() can be $|T|$, the number of
transitions in the TA. 
But it can be much tighter and more precise. 
In our implementation, we actually compute an untimed
quotient structure of $A$'s state-space through forward analysis
and eliminate those transitions that are actually not
triggerible. 
In this way, we usually come up with a much
smaller bound on $f$, which is the number of executable transitions in ACM.  

As for the computation of the numerical coverage estimation $v$,
we use $V$ as a static set variable of transitions such that
$V=\emptyset$ initially.
Each time when statement (4) in procedure {\tt Symbolic\_Simulate}() is executed,
we perform the following two steps.
\vspace*{3mm}
\hrule
\vspace*{2mm}
$V:= V\cup \bar{T}$; $v:=|V|$;
\vspace*{2mm}
\hrule
\vspace*{3mm}

{\lemma ACM for dense-time systems satisfies the accountability
criterion.}
\\\pf It is true since $\epsilon_{ACM}(e)=1$ for every executable transition in $F$, and 
we directly use the sizes of already-triggered and executable transition sets
to calcualte the coverage.  
\qed

The criterion of full coverability is not guaranteed.
But as can be seen from our experiment data in section~\ref{sec.experiments},
with a tight estimation of the set of executable transitions,
it is possible to get very close to 100\% of coverage.
As for the criterion of efficiency, in each iteration,
the overhead is a set-union operation and a size calculation of set 
and the efficiency is high.
Finally, ACM may not have much discernment
since a transition can very often be used in both a safe trace and a trace that ends
in a risk state. 
This means ACM may reach 100\% coverage without discovering the risk state even if it exists.

\section{Back-and-forth region coverage metric (RCM)\label{sec.rcm}}

ACM can very often be too coarse to discern risk states.
Another extreme that can also be adapted from 
VLSI verification technology is the {\em visited-state} coverage metric\cite{BF01,RPS01}, 
which uses the reachable state set in FSM to estimate coverage.
The challenge to incorporate the concept into our framework
arises from the fact that in VLSI's model, the states are discrete
and countable while in timed automata, the states are dense and uncountable.
A solution is to use equivalence classes in the dense-time state-space as portions.
An equivalence relation to partition dense-time state-space is the {\em region}-equivalence relation
between states\cite{ACD90}.
In this paper, a region is a {\em minimal} state-space characterizable by
a mode and clock-difference
constraints in the form of $x-x'\sim c$ where
$x,x'$ are two dense-time clocks, $\sim\in\{<,\leq,=,\geq,>\}$,
and $c$ is an integer in the range of $[-C_{A:\eta},C_{A:\eta}]$
where $C_{A:\eta}$ is the biggest timing constant used in $A$ and the
safety state predicate $\eta$.
In this way, we consider in this section the concept of
{\em region coverage metric (RCM)}, in which a basic portion is a region,
for the simulation of real-time systems.
This coverage metric can have extra leverage with symbolic simulation
since, in one step, we may generate a huge proportion of the state-space represented
by a set of symbolic constraints.

There are three challenges in the implementation of RCM.
First, how do we construct a tight estimation relevant to the reachability of
the states?
Second, how do we compute the coverage estimations of sets of portions,
i.e. $\sum_{p\in V}\epsilon_{RCM}(p)$ and $\sum_{p\in F}\epsilon_{RCM}(p)$?
Third, how do we maintain the accountability of the metrics?
In this section, we counter these three challenges in three steps.
For the first challenge, we use the intersection of abstractions of both backward and forward
reachable state-space representations to construct a tight estimation
of the whole target function (i.e., the whole reachable region set in RCM).
For the second challenge, we work on the level of {\em zones}.
A {\em zone} is a set of regions whose state-spaces
are characterizable by a set of clock difference constraints.
We then develop a procedure to calculate the region coverage estimation of a zone in the state-space.
Our estimation is efficient, because we adopt the concept of region to partition the dense state-space and
use CDD(Clock-Difference Diagram)\cite{BLPWW99} as our data structure.
Many model-checkers for TAs are built around the central manipulation
procedures of zones\cite{Alur99,PL00,Wang00,Wang01,Wang03,Yovine97}.
Finally, for the third challenge, 
we present a data-structure and show that the data-structure can
represent a state-space as a set of disjointed zones.
With this data-structure, we can estimate the coverage of a state-space
as the sum of coverages of a set of disjointed zones.

\subsection{Tight estimation of the target function}

In general, it is very expensive to compute representations for the exact
reachable state-spaces.
In our previous implementations,
we use abstractions of either the backward or the forward reachable state-spaces
to compute the estimation for the whole target function in RCM.
But such estimations seem very imprecise.
In some experiments, the final coverage estimations, when the whole reachable state-space
representations have been constructed, fall in the range of $10^{-5}$.
Moreover, much proportion of the reachable state-space
seems irrelevant to the reachability from initial states to risk states.

We have observed that to analyze this reachability, we only have
to trace through those states which are both backward reachable
from a risk state and forward reachable from an initial state. So
we use the following steps to compute an estimation of the whole
target function.
\vspace*{3mm} \hrule \vspace*{2mm}
{\small
Compute $F$ as the untimed quotient structure
\hstt 	of the state-space of $A$ from initial states.
\hst Compute $B$ as the magnitude quotient structure
\hstt	of the backward reachable state-space of $A$
	from risk states in $\neg \eta$.
\hst Let $F:=F\wedge B$;
}
\vspace*{2mm} \hrule \vspace*{3mm}
In the second statement, we employ an abstraction technique, called {\em
magnitude abstraction}, to compute the weakest preconditions from a
state-predicate. A magnitude abstraction of a state-predicate
eliminates from the state-predicate all clock inequalities like
$x-x'\sim c$ where $x,x'$ are not zeros.

With these three steps, we have constrained $F$ to a much smaller
state-space that is relevant to the reachability from initial states to
risk states. 
Notice that these steps should not be regarded as extraneous expenses for RCM,
since our symbolic simulator will initially take these steps
to shrink the state-space that we have to search during the simulation anyway.
According to our experiments reported in section~\ref{sec.experiments},
this technique has brought our ultimate estimation in RCM very close to 100\% and resulted
in much better coverability.

\subsection{Coverage estimation of a zone \label{sec.rcm.zone}}

A {\em zone} is a state-space characterizable by
a range constraints on the {\tt mode} variable and
a set of range constraints on the clock differences.
Conveniently speaking, the characterization can be represented as 
a pair like $(Q',K)$ such that
\begin{list1}
\item $Q'\subseteq Q$ and is the range of the {\tt mode} variable; and
\item $K$ is a set of range constraints like $c\sim x-x'\sim' c'$
    for clock differences, where $\sim,\sim'\in\{<,\leq\}$ and
    $c,c'\in [-C_{A:\eta},C_{A:\eta}]\cup\{-\infty,\infty\}$.
\end{list1}
For the efficiency of coverage estimation, we intuitively compute something like a {\em normalized
volume} estimation of zones.  
The volume estimation of a
rectangular polyhedron in a multi-dimensional space can be computed as the
multiplication of its length in each dimension.  
For efficiency, we intuitively interpret a zone as a rectangular polyhedron in a space of
$1+|X|(1+|X|)$ dimensions. 
The range of variable {\tt mode}'s
value spans the first dimension while $x_i-x_j$, for each
$x_i,x_j\in X\cup\{0\}$ with $i<j$, spans a dimension. 
This intuitive and simplistic interpretation of zones neglects the fact
that constraints on clock differences are not independent of one
another.  
But our experiments show that
it helps us design an efficient and coverable metric for region coverage.

In measuring the length of a clock difference constraint,
we partition the real number lines into the following $4C_{A:\eta}+3$ basic intervals
\begin{center}
\footnotesize
$(-\infty,-C_{A:\eta})[-C_{A:\eta},-C_{A:\eta}]
\ldots[-1,-1](-1,0)[0,0](0,1)[1,1](1,2)\ldots
[C_{A:\eta},C_{A:\eta}](C_{A:\eta},\infty)$
\end{center}
and
use the number of basic intervals covered by the clock difference constraint
for the estimated length.
For example, $-3\leq x-x' < 2$ has length $10$ because it
covers $[-3,-3],(-3,-2),\ldots,[1,1],(1,2)$.

Such volume estimations can result in huge numbers not representable by
integers in computers' hardware.
Thus instead of using the absolute length in each dimension to compute the
estimated volume, we choose to use the normalized lengths (i.e.
the floating point numbers of the length divided by the maximum length of the difference
variables).
The normalized length for $Q'$ is thus $|Q'|/|Q|$.
The normalized length for a clock difference constraint is 
broken down to the following eight cases: 
\begin{list1} 
\item $(2(c'-c)+1)/(4C_{A:\eta}+3)$ 
	for $c\leq x-x'\leq c'$ 
	with $c\geq -C_{A:\eta}$ and $c'\leq C_{A:\eta}$,
\item $(2(c'-c))/(4C_{A:\eta}+3)$ 
	for either $c< x-x'\leq c'$ or $c\leq x-x'< c'$
	with $c\geq -C_{A:\eta}$ and $c'\leq C_{A:\eta}$,
\item $(2(c'-c)-1)/(4C_{A:\eta}+3)$ 
	for $c< x-x'< c'$ 
	with $c\geq -C_{A:\eta}$ and $c'\leq C_{A:\eta}$,
\item $(2(c'+C_{A:\eta})+2)/(4C_{A:\eta}+3)$ 
	for $-\infty< x-x'\leq c'$ 
	with $c'\leq C_{A:\eta}$,
\item $(2(c'+C_{A:\eta})+1)/(4C_{A:\eta}+3)$ 
	for $-\infty< x-x'< c'$ 
	with $c'\leq C_{A:\eta}$,
\item $(2(C_{A:\eta}-c)+2)/(4C_{A:\eta}+3)$ 
	for $c\leq x_1-x_1'<\infty$ 
	with $c\geq -C_{A:\eta}$,
\item $(2(C_{A:\eta}-c)+1)/(4C_{A:\eta}+3)$ 
	for $c< x_1-x_1'<\infty$ 
	with $c\geq -C_{A:\eta}$,
\item $1$ for $-\infty< x_1-x_1'<\infty$; /* this case is usually not represented in zones */
\end{list1} 
Accordingly, the estimated normalized volume of a zone $(Q',K)$ is
\begin{center}
$\frac{|Q'|}{|Q|}\cdot\Pi_{\mbox{\scriptsize "}c\sim x-x'\sim' c'\mbox{\scriptsize "}\in K}
(\mbox{the normalized length of }c\leq x-x'\leq c')$
\end{center}

\subsection{Coverage estimation as a set of disjointed zones}

\begin{figure*}[t]
\begin{center}
\input{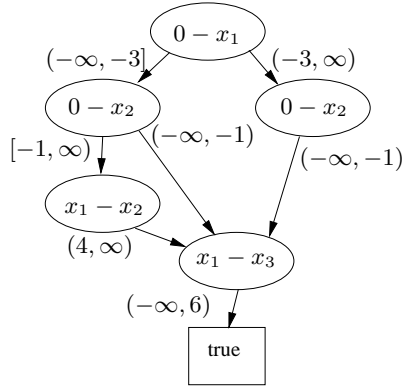}
\end{center}
\caption{\footnotesize CDD for $(0-x_1\leq -3\wedge x_1-x_3<-4\wedge x_2-x_1<6)
        \vee(0-x_2<-1\wedge x_2-x_1<6)$}
\label{fig.acdd}
\end{figure*}

Although the technique in the last subsection allows us to come up with
a coverage estimation of a zone,
the zones may intersect with one another and
accountability may not be maintained.
In this subsection, we present a representation for dense-time state-spaces
such that zones represented are disjointed from one another.
The representation that we have found with this property is
CDD\cite{BLPWW99} with all zones in their closure forms (or all-pair shortest-path form).
CDD is a BDD-like data-structure whose variables are clock differences like
$x-x'$ and whose outgoing arcs from variables are disjointed value ranges.
For example, the CDD for the state-space of
$(0-x_1\leq -3\wedge x_1-x_3<-4\wedge x_2-x_1<6)
\vee(0-x_2<-1\wedge x_2-x_1<6)$ without terminal $\false$ 
is in figure~\ref{fig.acdd}.
Each path in this figure represents a zone in closure form.
We refer interested readers to \cite{BLPWW99} for
the definition and manipulations of CDD.

We can prove the following lemma.

{\lemma \label{lemma.cov.region.disjointed}
Given a CDD with all zones in their closure forms,
then each two paths in the CDD represent two disjointed zones.}
\\\pf
From root to the terminals of the two paths,
there is a branching node from which the two paths break away.
The corresponding outgoing arcs from the node for the two paths
are labeled with disjointed intervals.
With the tightness of the zone constraints,
this means that the zones of the two paths are disjointed.
\qed

In our implementation, we follow the approach in \cite{BLPWW99} 
that a set of zones are first normalized to their 
closure forms before being stored in a CDD.  
Thus, for the convenience of presentation, we can 
assume that a CDD, say $D$, is represented as a $\true$, or a $\false$, 
or a tuple like $D=(x-x',(\lambda_1,D_1),\ldots,(\lambda_n,D_n))$, such that
\begin{list1}
\item the root node of $D$ is labeled with clock difference variable $x-x'$.
\item $\lambda_1,\ldots,\lambda_n$ are disjointed intervals whose endpoints
    are in $\{-\infty,-C_{A:\eta},\ldots,-1,0,1,\ldots,C_{A:\eta},\infty\}$.
\item for each $1\leq i\leq n$, the arc labeled with interval $\lambda_i$ points
    to $D_i$.
\end{list1}
With the desirable feature of CDD, we can design the following symbolic procedure
to compute the estimated volume of a state-space represented by
CDD with all zones in closure forms.
{\small 
\vspace*{3mm}
\hrule
\vspace*{2mm}
\noindent
{\tt normalized\_volume}$(D)$ 
/* $D$ is $\true$, $\false$, or $(x-x',(\lambda_1,D_1),\ldots,(\lambda_n,D_n))$ */ \{
\hst if $D$ is $\true$, return 1; else if $D$ is $\false$, return 0;
\hst else if $D$ is $(x-x',(\lambda_1,D_1),\ldots,(\lambda_n,D_n))$, then
\hstt   return $\sum_{1\leq i\leq n}
        \frac{\mbox{the number of basic intervals covered by }\lambda_i}
        {4C_{A:\eta}+3}
        \mbox{\tt normalized\_volume}(D_i)$;
\\\}
\vspace*{2mm}
\hrule
\vspace*{3mm}
}
Another advantage of this symbolic procedure is that we can take advantage 
of the data-sharing in CDD to avoid explicit enumeration of all disjointed zones.  
The normalized volume estimation of a substructure in a CDD can be saved 
and used for the estimation of other zones that use this same substructure.

\subsection{Estimation of the region coverage}

In our framework, both $V$ and $F$ in {\tt Symbolic\_Simulate}() 
on page~\pageref{proc.ss} are conceptually represented as a set of pairs like
$(Q',D)$, for a state-space, where
$D$ is a CDD with all zones in closure form, with the following constraints.
\begin{list1}
\item For each two pairs $(Q_1',D_1), (Q_2',D_2)$ in the set, $Q_1'\cap Q'_2=\emptyset$.
\item For each pair $(Q',D)$ in the set, $D$ represents the zones of all states with their modes
    in $Q'$.
\end{list1}
The procedure to transform a state-space representation in BDDs and DBMs
to this representation can be found in \cite{BLPWW99}.
Then at statement (4) of each iteration of procedure {\tt Symbolic\_Simulate}()
on page~\pageref{proc.ss},
the estimated normalized volume $v$ for $V$ is
$\sum_{(Q',D)\in V}\frac{|Q'|}{|Q|}\cdot \mbox{\tt normalized\_volume}(D)$
and
\linebreak
$\frac{v}{f}=\frac{\sum_{(Q',D)\in V}\frac{|Q'|}{|Q|}\cdot \mbox{\tt normalized\_volume}(D)}
		{\sum_{(Q',D)\in F}\frac{|Q'|}{|Q|}\cdot \mbox{\tt normalized\_volume}(D)}$.

{\lemma \label{lemma.rcm.accountability}
RCM satisfies the criterion of accountability.}
\\\pf
According to lemma~\ref{lemma.cov.region.disjointed}, zones respectively represented by paths in a CDD in its closure form
are disjointed from one another.
Thus in the algorithm of {\tt normalized\_volume}(), we count each portion once and only once
and RCM satisfies the criterion of accountability.
\qed


{\lemma \label{lemma.rcm.discern}
RCM satisifies the criterion of discernment, and it is impossible
to reach 100\% coverage without detecting the risk state, if any.
}
\\\pf
Given a timed automaton $A$ and a risk predicate $\eta$,
in RCM, safe states and unsafe states are not in the same portion.
This is because symbolic manipulations of zones are sufficient to
answer the reachability problem of timed automata\cite{ACD90}.
\qed

According to lemma~\ref{lemma.rcm.discern}, RCM has enough discerning power to
discover reachable risk states whereas ACM lacks such discerning power.

\section{Triggering-condition coverage metric (TCM) \label{sec.tcm}}

RCM has the advantage in accountability and discernment.
But it may result in low coverability since our estimation of the reachable region sets can 
be imprecise.
On the other hand, ACM can suffer from low discernment.
In this section, we propose a balanced approach called
{\em triggering-condition coverage metric (TCM)}, in which
we use the triggering conditions of all transitions as the body of
the whole target function.
TCM estimates the proportion of the covered triggering conditions of all transitions.
It is accounted as the summation of
triggering condition coverage for each transition.
Formally speaking, a basic portion in TCM is a pair 
like $(e,\gamma)$ where 
$e$ is an executable transition  
and $\gamma$ is a region in 
subspace $\tau(e)$ (the triggering condition of $e$).

The numerical estimation $f$ of the whole target function in procedure {\tt Symbolic\_Simulate}() can be computed as
$\sum_{e\in T}\mbox{\tt normalized\_volume}(\tau(e))$,
where $\tau(e)$ is the triggering condition of each $e\in T$.
We use $|T|$ variables, $V_e$ for each $e\in T$,
to record the verified proportion of the triggering condition of each transition.
Initially, for all $e\in T$, $V_e=\false$.
At each iteration's execution of statement (4),
we execute the following steps to compute the value $v$.
\vspace*{3mm}
\hrule
\vspace*{2mm}
for each $e\in \bar{T}$, $V_e:= V_e\vee (\mbox{\tt abstract}_e(\phi\wedge\tau(e)))$;
\hst let $v:=\sum_{e\in T}\mbox{\tt normalized\_volume}(V_e)$;
\vspace*{2mm}
\hrule
\vspace*{3mm}
Here for the sake of efficiency, we use an abstract function $\mbox{\tt abstract}_e(d)$ 
to eliminate the recording of all clock difference variables not used in $\tau(e)$.
For example, if $\tau(e)=0-x<-5\wedge x-y\leq 3$, then
$\mbox{\tt abstract}_e(0-x < -7 \wedge x-y \leq -2 \wedge y-0\leq 2)=
0-x<-7\wedge x-y\leq -2$ and the constraint literal $y-0\leq 2$ is filtered out
since no constraint on difference $y-0$ is used in the triggering condition of $e$.
Also in these two steps, we assume that while invoking
$\mbox{\tt normalized\_volume}(V_e)$ for each $e$,
$V_e$ has already been transformed to the representation like
$\{(Q'_1,D_1),\ldots,(Q'_n,D_n)\}$.

It can be shown that TCM has the following desirable property.

{\lemma TCM satisfies
the criterion of accountability.}
\\\pf We calculate the normalized volume of zones based on the
triggering conditions of transitions with TCM.
Since zones of the triggering conditions of each transition represented by a CDD
are disjointed, TCM satisfies the criterion of accountability.
\qed

TCM is more efficient than the RCM since
it is based on abstraction of zones whose representation complexity is usually
lower.
In the following experiments,
we shall see that it satisfies the criterion of coverability without
sacrificing  its discernment.

\section{Experiments with Bluetooth L2CAP \label{sec.experiments}}

We have implemented our numerical coverage estimation techniques
in our model-checker/simulator {\tt red} 4.1.
The input language of {\tt red} is a set of {\em communicating timed automata (CTA)}
that communicate with one another through CSP-style
synchronization channels\cite{Hoare85}. For each channel $\sigma$,
two processes have to execute at the same instant to achieve a
synchronization through the channel. In the synchronization
instant, one process executes a transition with event $!\sigma$
for output and the other executes a transition with event
$?\sigma$ for input.

To check the possibility of using our techniques in real-world projects,
we have experimented with the {\em
Logical Link Control and Adaptation Layer Protocol(L2CAP)} of
Bluetooth specification\cite{Haartsen01}.
The wireless communication standard of Bluetooth has been widely
discussed and adopted in many appliances since it was published.
L2CAP is layered over the Baseband Protocol and resides in the data link layer
of Bluetooth.
This protocol supports message multiplexing, packet
segmentation and reassembly, and the conveying of quality of
service information to the upper protocol layer.
The protocol regulates the behavior between a master device and a slave device.

In our experiment, we collect coverage and performance data for
L2CAP models both with and without design faults against various trace-generation strategies.
In subsection~\ref{subsec.nofaults}, we report the coverage data of
ACM, RCM, and TCM for the L2CAP model without faults.
In subsection~\ref{subsec.faults}, we create six versions of
the L2CAP model, each with an inserted fault, and report how
the coverage metrics help us discern the faults before 100\% coverage is reached.
In appendix~\ref{app.strategies}, more coverage and performance data of our experiments
with various trace-generation strategies can be found.
Data is collected on a Pentium 4 with 1.7GHz,
256MB, running Red Hat Linux 7.0.

\subsection{Modelling L2CAP \label{subsec.l2cap}}

The L2CAP defines the actions performed by a master and a slave.
A master is a device issuing a request while a slave is the one responding to
the master's request.
A message sequence chart (MSC) that may
better illustrate a typical scenario of event sequence in L2CAP
can be found in appendix~\ref{app.bmsc}.
The scenario starts when the master's upper layer
issues an {\tt L2CA\_ConnectReq} (Connection Request)
through the L2CA interface.
Upon receiving the request,
the master communicates the request through the unreliable network to the slave (with an {\tt L2CAP\_ConnectReq}),
which will then convey the request to the slave's upper layer
(with an {\tt L2CA\_ConnectInd}).

The protocol goes on with messages bouncing back and forth
until the master sends an {\tt L2CAP\_ConfigRsp} message to the slave.
Then both parties can start exchanging data.
Finally the master's upper layer issues message {\tt L2CA\_DisconnectReq} to
close the connection and the slave confirms the disconnection.


We use nine processes to model the entire activity in L2CAP.
They are
 the master's upper layer,
the master's L2CAP layer,
master's L2CAP time-out process,
master's L2CAP extended time-out process,
the slave's upper layer,
the slave's L2CAP layer,
slave's L2CAP time-out process,
slave's L2CAP extended time-out process,
and the unreliable network.
Each of these processes is described as a communicating timed automaton.
The CTA for both the master and the slave can be found in appendix~\ref{app.bdevice}.
The safety condition is that when the master's L2CAP layer stays in the OPEN state,
 the slave's L2CAP layer can not enter the state
W4\_L2CA\_DISCONNECT\_RSP.

\subsection{Coverage estimation when there is no fault  \label{subsec.nofaults}}

In this subsection, we execute procedure {\tt Symbolic\_Simulate}()
with {\em breadth-first} strategy to verify our L2CAP model without faults.
That is, each time we execute statement (2) in procedure {\tt Symbolic\_Simulate}(),
we let $\bar{T}=T$ and $\phi=\psi$.

In each iteration, we calculate three estimations according to the three coverage metrics
respectively.
The data is in table~\ref{tab.covtime.exp}.
\begin{table*}[t]
\begin{center}
\small
\begin{tabular}{|c||c|c||c|c||c|c|} \hline
iteration &  ACM & ACM time & RCM & RCM time & TCM & TCM time \\
	&  	& overhead & 	& overhead &	 &  overhead \\ \hline \hline
 1	& 4/97	& 0.00sec.	& 0.167816	& 7.39sec. & 0.004092	& 0.02sec. \\
 2	& 8/97	& 0.00sec.	& 0.173442	& 7.39sec.	& 0.382901	& 0.02sec. \\
 3	& 12/97 & 0.00sec.	& 0.174279	& 7.40sec.	& 0.783131	& 0.02sec. \\
 4	& 20/97 & 0.01sec. & 0.175273	& 7.41sec.	& 0.799498	& 0.02sec. \\
 5	& 36/97 & 0.02sec. & 0.232154	& 7.41sec.	& 0.813138	& 0.03sec. \\
 6	& 42/97 & 0.03sec. & 0.295386	& 7.41sec.	& 0.815525	& 0.04sec. \\ 
 7	& 64/97 & 0.05sec. & 0.408160	& 7.42sec.	& 0.884971	& 0.06sec. \\
 8	& 76/97 & 0.08sec. & 0.561395	& 7.43sec.	& 0.920890	& 0.08sec. \\
 9	& 88/97 & 0.12sec. & 0.956820	& 7.44sec.	& 0.971241	& 0.11sec. \\
 10	& 94/97 & 0.17sec. & 0.965724	& 7.45sec.	& 0.975507	& 0.15sec. \\
 11	& 94/97 & 0.22sec. & 0.974428	& 7.46sec.	& 0.975507	& 0.18sec. \\
 12	& 95/97 & 0.28sec. & 0.975538	& 7.48sec.	& 0.976530	& 0.22sec.	\\ 
 13	& 97/97 & 0.34sec. & 0.975783	& 7.49sec.	& 1.000000	& 0.26sec. \\
 14	& 97/97 & 0.40sec. & 0.981319	& 7.50sec.	& 1.000000	& 0.29sec. \\
 15	& 97/97 & 0.47sec. & 0.981338	& 7.52sec.	& 1.000000	& 0.33sec. \\
 16	& 97/97 & 0.55sec. & 0.982733	& 7.54sec.	& 1.000000	& 0.36sec. \\
 17	& 97/97 & 0.63sec. & 0.982734	& 7.56sec.	& 1.000000	& 0.40sec. \\
 18	& 97/97 & 0.70sec. & 0.982734	& 7.57sec.	& 1.000000	& 0.44sec. \\  \hline
\end{tabular}
\caption{Coverage estimations and overheads with respect to iterations when there are no bugs}
\end{center}
\label{tab.covtime.exp}
\end{table*}
After 18 iterations, {\tt red} 4.1 finishes the exhaustive search,
and reports that the risk state is NOT reachable.
It costs total cpu time 37.14 sec and memory usage 782k with ACM;
total cpu time 30.59 sec and memory usage 632k with RCM;
total CPU time 35.79 sec and memory usage 722k with TCM.

ACM and TCM can both reach 100\% coverage estimation
while RCM gets very close to 100\%.
The data shows that our methods have very high coverability in the experiment.

Another interesting thing is that for this correct L2CAP model,
ACM and TCM can give us 100\% confidence in their respective
metrics before the whole reachable state-space representation is constructed.
More precisely,
according to ACM and TCM, we can stop at iteration 13 with 100\% confidence.
On the other hand, if we use straightforward formal verification,
then we have to run through
all the 18 iterations before we can conclude that the model is fault-free.
This observation suggests that symbolic simulation with our coverage metrics can
greatly save verification costs.

Since RCM gets us very close to 100\% coverage,
we can use 100\% coverage as a goal for verification in RCM.
More importantly, RCM is a better alternative in discernment than ACM and TCM.
For one thing, at the 17th iteration, it could still increase to reflect
more portions that have been traced through while ACM and TCM have already converged to 1.

\begin{figure}[t]
\begin{center}
\epsfig{height=90mm, file=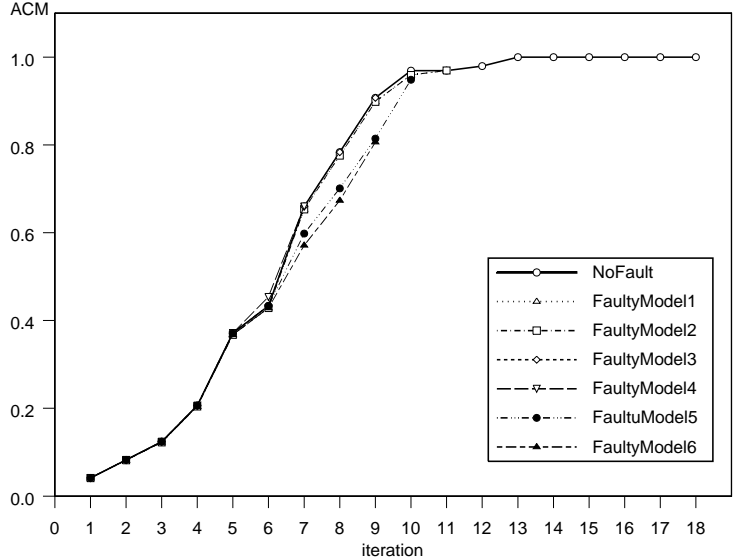}
\end{center}
\caption{Growth of coverage with respect to iterations in ACM} \label{figure.fault.acm.exp}
\end{figure}

\begin{figure}[t]
\begin{center}
\epsfig{height=90mm, file=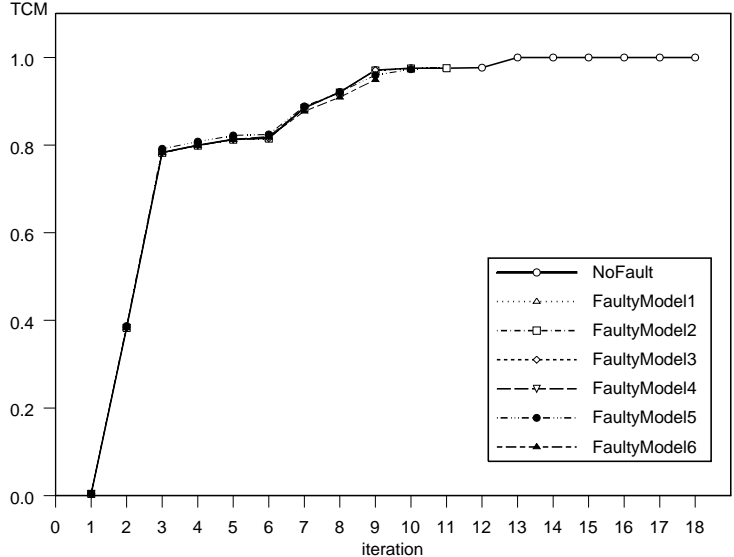}
\end{center}
\caption{Growth of coverage with respect to iterations in TCM} \label{figure.fault.tcm.exp}
\end{figure}

As for the efficiency of our coverage estimation methods,
we find that at the end of the state-space construction,
the overhead incurred in the coverage estimation respectively for ACM, RCM and TCM  is about 0.70, 7.57 and 0.44 seconds.
Compared with the verification time,
we find that for ACM, only $0.70/37.14\approx 0.01885=1.885\%$ of the
verification CPU time is used in the coverage estimation.
For TCM, only $0.44/35.79\approx 0.01229=1.229\%$ of the
verification CPU time is used.
This means that our implementation for both ACM and TCM are quite efficient.
In figure~\ref{figure.fault.acm.exp} and figure~\ref{figure.fault.tcm.exp},
we drew the growth of coverage in ACM and TCM for the correct model and the six faulty models (details in the next subsection) with respect to the iterations.
Notice that both coverage metrics grow quickly between the 4th iteration and the 10th iteration
and then become flattened out to convergy to 100\%.
It reaches 100\% at the 13th iteration and finishes the exhaustive search at the 18th iteration in the correct case.
In the all faulty models we reach the risk state before the 11th iteration.
The patterns show that both metrics may give engineers enough confidence to make decision quickly. 
For example, they may stop the simulation while the coverage becomes 100\% or whilr it starts to converge 100\%,
and save the verification resources.

\begin{figure}[t]
\begin{center}
\epsfig{height=90mm, file=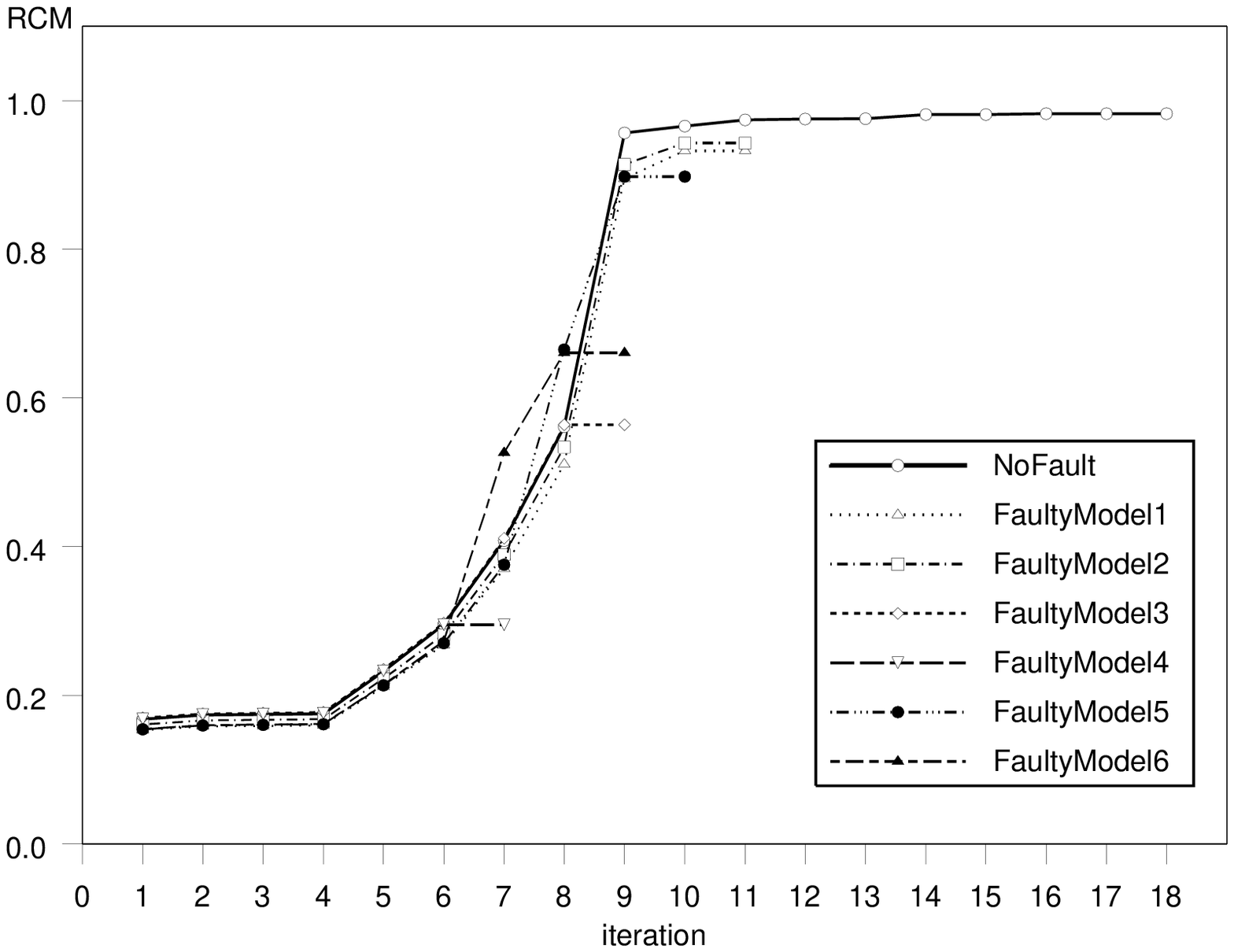}
\end{center}
\caption{Growth of coverage with respect to iterations in RCM} \label{figure.fault.RCM.exp}
\end{figure}

For RCM, $7.57/30.59\approx 0.24747=24.747\%$ of the verification CPU time 
is used in the coverage estimation.  
A detail breakdown of the computation time shows that most of the overhead is
consumed in the normalized volume calculation with the much larger CDD structure.
This is the price of better discernment.  

Figures~\ref{figure.fault.RCM.exp} shows the growth of coverage in RCM for the same models.
It's interesting that the patterns dramatically increase in the first few iterations and 
then slow down in the next ierations.
We can also detect the faults before the 11th iteration while the coverage in the correct model stops increasing in the 17th iteration
and finishes the exhaustive search at iteration 18.
Although RCM could not satisfy the coverability, we can figure out the end point while it stops increasing.

\subsection{Coverage estimation when there is a fault \label{subsec.faults}}

We design six L2CAP faulty models, each with an inserted fault.
For convenience, we label these six faulty models with indices 1 through 6.
In each faulty model, we change master or slave's behaviors and let the risk condition
become reachable.
The description of the six faulty models are given in appendix~\ref{app.faulty.models}.  
We tried two trace-generation strategies.
The first is breadth-first (see subsection~\ref{subsec.nofaults}).
The second is {\em depth-first}.
That is, at each time when we execute statement (2), we choose $\bar{T}$ to be of size 1 and
only choose to fire one transition in $T$.
We also keep a stack so that we can backtrack to previous iterations to choose an
alternative transitions at statement (2).
The coverage data is shown in table ~\ref{tab.forward.exp}.
\begin{table*}[t]
\begin{center}
\small
\begin{tabular}{|c|c|r|c|c|c|c|} \hline
Strategy & Faulty Models & Depth  & ACM   & RCM   & TCM   & Risk state reached?\\ \hline \hline
Depth	& 1	& 66	& 71/98 	& 0.355262	& 0.958950	& Yes	\\
First	& 2	& 64	& 71/98	& 0.354993	& 0.958950	& Yes	\\
	& 3	& 26	& 27/97	& 0.293884	& 0.437878	& Yes	\\
	& 4	& 96	& 90/97	& 0.924978	& 0.963982	& Yes	\\
	& 5	& 64	& 63/97	& 0.355688	& 0.948209	& Yes	\\
	& 6	& 62	& 61/98	& 0.357885	& 0.936412	& Yes	\\ \hline
Breath	& 1	& 11	& 95/98	& 0.932724	& 0.975532	& Yes	\\
First	& 2	& 11	& 95/98	& 0.943274	& 0.975532	& Yes	\\
	& 3	& 9	& 88/97	& 0.564228	& 0.971241	& Yes	\\
	& 4	& 7	& 64/97	& 0.294859	& 0.884971	& Yes	\\
	& 5	& 10	& 92/97	& 0.898077	& 0.973172	& Yes	\\
	& 6	& 9	& 79/98	& 0.660754	& 0.949243	& Yes	\\ \hline
\end{tabular}
\end{center}
\caption{Coverage estimations with respect to two strategies for the 6 faulty models}
\label{tab.forward.exp}
\end{table*}
The most interesting thing in table~\ref{tab.forward.exp} is that
the faults are all detected before we reach 100\% coverage.
This means that our three coverage metrics have enough discernment for
the six faulty models.

\section{Conclusion}

Symbolic simulation combines the advantages of both simulation and formal
verification and can be an important verification approach before fully automatic
formal verification becomes applicable.
In this paper, we present techniques for coverage estimation for dense-time systems.
We hope such techniques can be the solid stepstone toward the development of
powerful symbolic simulators for industry real-time systems.  
Many issues raised in this work also deserve future research.  
For example, it will be interesting to see the design of quantitative 
metrics for our criterion of discernment in the symbolic simulation of 
dense-time systems.  
With such metrics, the criterion becomes equivalent to the notion of 
observability\cite{DGK96}.

\newpage
\appendix
\noindent{\bf \LARGE APPENDICES} 
\pagenumbering{roman}
\setcounter{page}{1}
\setcounter{section}{0}

\section{Message sequence chart for Bluetooth L2CAP\label{app.bmsc}}

\begin{figure}[h]
\begin{center}
\input{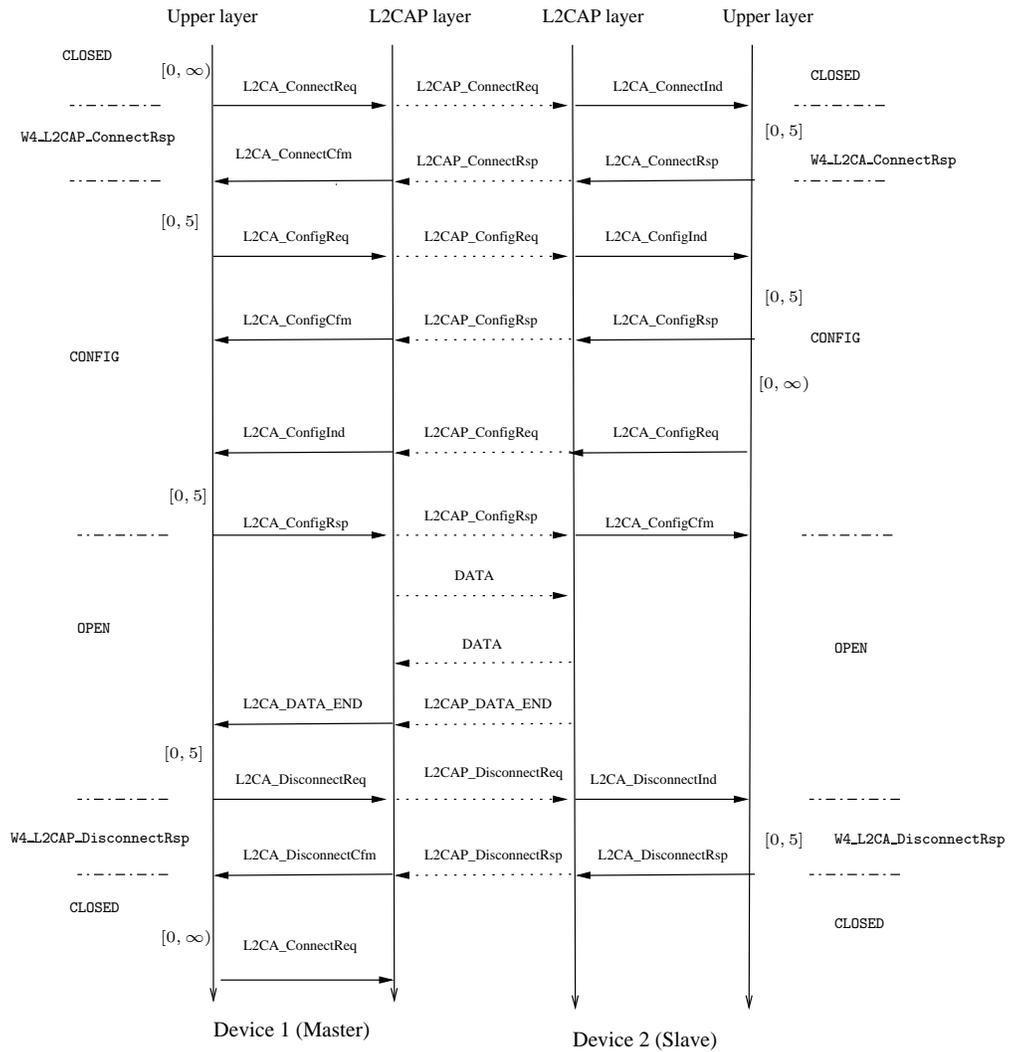}
\end{center}
\caption{A message sequence chart of L2CAP}
\label{fig.bmsc}
\end{figure}
The two outer descriptions represent the states of L2CAP layers.

\newpage
\section{Bluetooth L2CAP \label{app.bdevice}}

\begin{figure*}[h]
\hspace*{-2cm}
\begin{center}
\input{bl2cap.tex}
\end{center}
\caption{CTA of a Bluetooth device}
\label{fig.bdevice}
\end{figure*}

\section{Description of the six faulty models\label{app.faulty.models}}

All faulty models lead the risk state reachable and violate the safety condition.
Recall that the safety condition is that while master L2CAP device still stays in the OPEN state
at the time, the slave L2CAP device may not enter the
W4\_L2CA\_DISCONNECT\_RSP state.
We compare the fault and correct behaviors of these faulty models as below:

\begin{list1}
\item $\mbox{\tt Faulty Model1}:$ 
The slave will enter into W4\_L2CA\_DISCONNECT\_RSP state while receiveing master's data from network and notifying upper layer.
In the correct model, the slave shall stay in OPEN state.
\item $\mbox{\tt Faulty Model2}:$ 
The slave will leave OPEN state and enter into W4\_L2CA\_DISCONNECT\_RSP state 
while reciveing upper layer's disconnect command, sending this request to master through network, and starting the timer.
In the correct model, the slave shall leave OPEN state and enter into W4\_L2CAP\_DISCONNECT\_RSP state 
to wait for the master's response from network.
\item $\mbox{\tt Faulty Model3}:$
The master remains staying in OPEN state while reciveing upper layer's disconnect command, sending this request to slave through network, and starting the timer.
In the correct model, the master shall leave OPEN state and enter into W4\_L2CAP\_DISCONNECT\_RSP state 
to wait for the slave's response from network.
\item $\mbox{\tt Faulty Model4}:$ 
The master will leave CONFIG state and enter into OPEN state while reciveing upper layer's disconnect command, sending this request to slave through network, and starting the timer.
In the correct model, the master shall enter into W4\_L2CAP\_DISCONNECT\_RSP state 
to wait for the slave's response from network.
\item $\mbox{\tt Faulty Model5}:$ 
The slave will leave CONFIG state and enter into W4\_L2CAP\_DISCONNECT\_RSP state while receiving upper layer's configuration response and having received master's response.
In the correct mode, the slave shall leave CONFIG state and enter into OPEN state after finishing the configuration process.
\item $\mbox{\tt Faulty Model6}:$  
The slave will leave CONFIG state and enter into W4\_L2CAP\_DISCONNECT\_RSP state while receiving upper layer's configuration response but not yet having received master's response.
In the correct mode, the slave should stay in CONFIG state since it doesn't finish the configuration process.
\end{list1}

\section{Coverage estimation with various search strategies
\label{app.strategies}}

It is also interesting to see how our techniques can be used
together with various trace-generation strategies in symbolic simulation.
We only briefly describe the trace-generation strategies
that we have implemented for our experiments in the following.
\begin{list1}
\item {\em Random Walk Strategy}:
    Each time we execute statement (2), {\tt red} 4.1 randomly
    choose a firable transition to be the sole element in $\bar{T}$.
\item {\em Game-based Strategy}:
    We use the term {\em "game"} here because we envision the concurrent system
    operation as a game.
    Users can specify some processes to be treated as {\em players}
    while the other processes are treated as {\em opponents}.
    At each time we execute statement (2),
    we either randomly choose a firable transition from the opponent processes
    or choose $\bar{T}$ to be the set of all firable transitions of the player processes.
    In this strategy, we alternately execute sequences of all players' transitions and
    sequences of a random-walk of the opponents' transitons.
    In this experiment, we view all processes whose local variables and clocks appear
    in the safety predicate as players.
    All other processes are opponents.
\item {\tt Goal-oriented Strategy}:
    According to this strategy, heuristics are designed for the choice of
    a single transition in each execution of statement (2) in the
    hope that a short trace leading to a risk state can be constructed.
\end{list1}
Our coverage data with the generation of a single symbolic trace is in table~\ref{tab.strategy.exp}.
\begin{table*}[t]
\begin{center}
\small
\begin{tabular}{|c|c|c|c|c|c|c|} \hline
Strategy & Faulty Models & Depth  & ACM   & RCM   & TCM   & Risk state reached?\\ \hline \hline
Random	& 1	& 21	& 34/98	& 0.437014	& 0.623092	& No	\\
Walk	& 2	& 11	& 23/98	& 0.181328	& 0.423497	& No	\\
	& 3	& 11       & 22/97	& 0.172719	& 0.411343	& No	\\
	& 4	& 8	& 26/97	& 0.168460	& 0.217365	& No	\\
	& 5	& 30	& 44/97	& 0.294042	& 0.839828	& No	\\
	& 6	& 19	& 27/98	& 0.179093	& 0.636663	& No	\\ \hline
Goal	& 1	& 15	& 19/98	& 0.187034	& 0.613695	& No	\\
Oriented & 2	& 14	& 12/98	& 0.164324	& 0.597204	& No	\\
	& 3	& 8	& 26/97	& 0.169848	& 0.217365	& No	\\
	& 4	& 13	& 12/97	& 0.199808	& 0.597815	& No	\\
	& 5	& 14	& 28/97	& 0.157394	& 0.613456	& No	\\
	& 6	& 19	& 19/98	& 0.163539	& 0.613695	& No	\\ \hline
Game	& 1	& 14	& 30/98	& 0.429387	& 0.421141	& No	\\
Based	& 2	& 8	& 26/98	& 0.160757	& 0.217143	& No	\\
	& 3	& 23	& 33/97	& 0.309106	& 0.631373	& No	\\
	& 4	& 17	& 29/97	& 0.188423	& 0.429172	& No	\\
	& 5	& 20	& 32/97	& 0.183961	& 0.626350	& No	\\
	& 6	& 24	& 20/98	& 0.158451	& 0.798677	& No	\\ \hline
\end{tabular}
\end{center}
\caption{Coverage estimations with respect to automatic trace generation
strategies for the 6 faulty models} \label{tab.strategy.exp}
\end{table*}
The experiments shows that it is viable to integrate our techniques with
other verification and simulation techniques.

\end{document}